\documentclass[12pt]{iopart}
\usepackage{iopams}  
\usepackage{cite}
\usepackage[english]{babel}

\def\C{\mathbb{C}}

\def\N{\mathbb{N}}

\newcounter{theo}
\newcommand{\theo}{\refstepcounter{theo}\noindent\textbf{Theorem \thetheo.} }
\newcounter{prop}
\newcommand{\prop}{\refstepcounter{prop}\noindent\textbf{Proposition \theprop.} }
\newcounter{rem}
\newcommand{\rem}{\refstepcounter{rem}\noindent\textbf{Remark \therem.} }
\newcounter{exam}
\newcommand{\exam}{\refstepcounter{exam}\noindent\textbf{Example \theexam.} }
\newcounter{lem}
\newcommand{\lem}{\refstepcounter{lem}\noindent\textbf{Lemma \thelem.} }
\newcounter{defi}
\newcommand{\defi}{\refstepcounter{defi}\noindent\textbf{Definition \thedefi.} }
\begin{document}

\title[Non-commutative generalization of integrable quadratic ODE systems]{Non-commutative generalization of integrable quadratic ODE systems}

\author{V.\ Sokolov $^{abc}$ and T.\ Wolf $^{def}$}

\address{$^a$ Landau Institute for Theoretical Physics, Chernogolovka, Russia\\
$^b$ UFABC, Sao Paulo, Brazil \\
$^c$ ORCID: 0000-0002-8937-1945 \\
$^d$ Department of Math \& Stat, Brock University, St Catharines, ON, Canada\\
$^e$ ORCID: 0000-0003-4136-8859 \\
$^f$ Tel: +1 905 688 5550 x 3803}
\ead{vsokolov@landau.ac.ru, twolf@brocku.ca}
\begin{abstract}
We find all homogeneous quadratic systems of ODEs with two dependent variables that have polynomial first integrals and satisfy the Kowalevski-Lyapunov test. Such systems have infinitely many polynomial infinitesimal symmetries. We describe all possible non-commutative generalizations of these systems and their symmetries.  
As a result, new integrable quadratic homogeneous systems of ODEs with two non-commutative variables are constructed. Their integrable non-commutative inhomogeneous generalizations are found. In particular, a non-commutative generalization of a Hamiltonian flow on the elliptic curve is presented. \vspace{12pt} \vspace{12pt}  
\end{abstract}

\section{Introduction}

We consider the   classification problem for integrable systems of the form
\begin{equation}
\left\{
\begin{array}{lcl}
u_t   =  \alpha_1 u^2 + \alpha_2 u \, v + \alpha_3  v \, u +
\alpha_4 v^2\\[2mm]
v_t   =  \beta_1 v^2  + \beta_2 v  \, u + \beta_3  u \, v + \beta_4
u^2 \label{equ}
\end{array}  , 
\right.
\end{equation}
where $u$ and $v$ are non-commutative symbols, which generate  the free associative algebra
 ${\mathfrak A}$ over  $\C$. Alternatively one may regard $u(t)$ and $v(t)$  as matrices of arbitrary size $m\times m$. By integrability we mean the existence of an infinite hierarchy of infinitesimal  polynomial symmetries. 
 
 Some partial results in this direction were obtained in \cite{miksokcmp}. We get the complete classification under two natural assumptions:
 \begin{itemize}
 \item[{\bf 1.}] The system (\ref{equ}) is integrable in a common sense if the size of matrices is equal to 1. 
 \end{itemize}
In the case $m=1$ we arrive at the usual system of ODEs of the form 
\begin{equation}\label{absys}
\left\{
\begin{array}{lcl}
u_t &=& a_1 u^2 + a_2 u v + a_3 v^2, \\[2mm]
v_t &=& b_1 v^2 + b_2  u v + b_3 u^2 
\end{array}
\right.
\end{equation}
where 
$$
a_1=\alpha_1, \quad a_2=\alpha_2+\alpha_3, \quad a_3=\alpha_4, \qquad b_1=\beta_1, \quad b_2=\beta_2+\beta_3, \quad b_3=\beta_4.
$$
The system (\ref{absys}) is called the {\it commutative limit} of the system (\ref{equ}). 

\noindent Moreover we assume that 
\begin{itemize}
 \item[{\bf 2.}] the commutative limit of the symmetry hierarchy of system (\ref{equ}) coincides with the whole symmetry hierarchy of (\ref{absys}).
\end{itemize}

In Section \ref{homabelian} we describe all integrable ODE systems of the form (\ref{absys})  up to linear transformations 
\begin{equation}\label{invert}
\hat{u}=r_1 u+r_2 v\, ,\qquad \hat{v}=r_3 u+r_4 v\, ,\qquad r_i \in \C, 
\qquad 
r_1 r_4 -r_2 r_3 \ne 0.
\end{equation}
Two systems related by such a transformation are called {\it equivalent}.

In Section \ref{homnonabelian} for each integrable system (\ref{absys}) we find all non-triangular integrable {\it non-commutative generalizations} i.e.  systems (\ref{equ}) such that (\ref{absys}) is their commutative limit.

Section \ref{inhomgen} is devoted to non-homogeneous generalizations of the form 
\begin{equation}
\left\{
\begin{array}{lcl}
u_t   =  \alpha_1 u^2 + \alpha_2 u \, v + \alpha_3  v \, u +
\alpha_4 v^2 + \gamma_1 u +\gamma_2 v+\gamma_3\, {\rm I},\\[2mm]
v_t   = \beta_1 v^2  + \beta_2 v  \, u + \beta_3  u \, v + \beta_4
u^2  + \gamma_4 u + \gamma_5 v + \gamma_6\, {\rm I}\label{inhom}
\end{array} ,  
\right.
\end{equation}
of integrable systems (\ref{equ}) from Section \ref{homnonabelian}. The system (\ref{inhom}) coincides with (\ref{equ}) up to terms of lower degree. The system (\ref{equ}) is called the {\it homogeneous limit} of a system (\ref{inhom}). We find all such integrable non-homogeneous generalizations.

The most interesting  example (see \cite{makont}) of such non-homogeneous integrable systems is given by
\begin{equation}
\left\{
\begin{array}{lcl}
u_t   =  v^2 + c u + a \, {\rm I},\\[2mm]
v_t   =  u^2 - c v + b \, {\rm I}\label{inuv}
\end{array} .  
\right.
\end{equation}
In the matrix case $u$ and $v$ are $m\times m$-matrices, ${\rm I}$ is the identity matrix,
 $a , b$ and $c$ are arbitrary scalar constants. This system turns out to be Hamiltonian. In Section \ref{sub6.1} we prove  the
integrability of (\ref{inuv})  by finding a Lax representation for the system. 
 
 The existence of a Lax pair allows one to integrate the system by the inverse scattering method  (see, for example, \cite{AblSeg81}). The  AKS-scheme \cite{AKS}) is another method for integration. Sometimes it is possible to find an explicit transformation that reduces the non-linear system to a linear one. 
 
 We are sure that any matrix system found in this paper can be integrated by one of these methods. Some examples are given in Section \ref{section5} but a detailed investigation of each system is a subject for a separate paper. 
 
 The  approach based on the assumptions {\bf 1} and {\bf 2} can be used for finding integrable non-commutative generalizations of diverse polynomial integrable models. 
 
\section{Integrable scalar homogeneous  polynomial systems with two variables} \label{homabelian}

Let us consider integrable systems of ODEs of the form 
\begin{equation}\label{quadNsc}
\frac{d y^i}{d t} = \sum_{j,k} C^i_{jk}\, y^j \,y^k, \qquad i,j,k=1,\dots, n,
\end{equation}
where $C^i_{jk}=C^i_{kj}$ are constants. The main feature of integrable systems of the form (\ref{quadNsc}) is the existence of infinitesimal polynomial symmetries and first integrals.

Another evidence of the integrability is the absence of movable singularities in solutions for complex $t$. The so-called Painlev\'e approach is based on this assumption. We use the {\it Kowalevski-Lyapunov test}, which is one of the incarnations of the Painlev\'e approach for polynomial homogeneous systems. 

\defi \label{defi1} The dynamical system
\begin{equation}
\frac{d\,y^{i}}{d\tau} = G_{i}(y^1,\ldots ,y^n),\ \ \ \ i=1, \ldots
,n  \,   \label{dynsym}
\end{equation}
is called {\it infinitesimal symmetry} for a dynamical system 
\begin{equation}
\frac{d\,y^{i}}{dt} = F_{i}(y^1,\ldots ,y^n),\ \ \ \ i=1, \ldots ,n
\,   \label{dynsys}
\end{equation}
iff (\ref{dynsys}) and (\ref{dynsym}) are compatible. The compatibility means that $$X Y-Y X=0,$$ where 
the vector fields $X$ and $Y$ are given by
$$\displaystyle X=\sum F_i \frac{\partial}{\partial y^i}, \qquad 
  Y=\sum G_i \frac{\partial}{\partial y^i}.$$

\defi \label{defi2} The function $I(y_1,\dots,y_n)$ is called {\it first integral}
for the system (\ref{dynsys}) iff
$$
X(I)=0.
$$

Systems of the form  (\ref{quadNsc}) possess special Kowalevski solutions of the form
$$
y^i(t)=\frac{b^i}{t},
$$
where
$$
-b^i = \sum_{j,k} C^i_{jk}\, b^j \,b^k.
$$
For any Kowalevski solution consider a formal solution of the form
$$
y^i= \frac{b^i}{t} + \varepsilon \, p^i \, t^{s}+ \cdots
$$
The numbers $p^i$ are defined from the system of linear equations
\begin{equation}\label{Kowsys}
s\, p^i = \sum_{j,k} C^i_{jk}\,(p^j b^k +\,p^k b^j), \qquad i=1,\dots, n.
\end{equation}
We see that the {\it  Kowalevski exponent} $s$ is an eigen-value and $(p^1,\dots p^n)$ is the corresponding eigen-vector of the linear system (\ref{Kowsys}).

\defi \label{defi3} System (\ref{quadNsc}) satisfies the  {\it Kowalevski-Lyapunov} integrability test iff
for any Kowalevski solution all exponents $s$ are integers.

In this section we find all systems of the form (\ref{absys})
that possess polynomial first integrals and satisfy the  Kowalevski-Lyapunov test.  

The following statement is well known and can be easily verified.

\lem \label{lem1} Let $I$ be a first integral of a system (\ref{absys}). Then for any  $N$
$$
\left\{
\begin{array}{lcl}
u_{\tau} &=& I^N\,(\alpha_1 u^2 + \alpha_2 u v + \alpha_3 v^2), \\[2mm]
v_{\tau} &=& I^N\, (\beta_1 v^2 + \beta_2  u v + \beta_3 u^2)
\end{array}
\right.
$$ 
is a (infinitesimal) symmetry of system (\ref{absys}).

It follows from this Lemma, that all systems we are going to find in this section  have an infinite hierarchy of homogeneous polynomial symmetries.

Let $I(u,v)$ be a homogeneous polynomial integral of a system (\ref{absys}).  
Let us write $I$ in the factorized form:
\begin{equation}\label{fint}
I = \prod_{i=1}^k (u-\kappa_i v)^{n_{i}}, \qquad n_i \in \N,  \qquad \kappa_i\ne \kappa_j \,\, {\rm if} \,\,i\ne j.
\end{equation}
\rem A factor of the form $v^{n}$ in $I$ corresponds to $\kappa =\infty.$ Using a linear transformation (\ref{invert}) of the form $\hat{u} = u,\, \hat{v} = v + c u$, one can always reduce $I$ to an integral of the form (\ref{fint}) with finite roots $\kappa_i, \, i = 1, \dots, k.$

\prop \label{prop4} Suppose that at least one of the coefficients of a system (\ref{absys}) is not equal to zero. Then $k\le 3. $ 

{\bf Proof}. Substituting $u=\kappa_i v$ for $i = 1, \dots ,k$ into the expression $Q=\frac{d I}{d t}$, we obtain that if $Q=0$ then each of roots $\kappa_i$ satisfies the same cubic equation
$$
b_3 \kappa_i^3+(b_2 -a_1) \,\kappa_i^2+(b_1-a_2)\, \kappa_i +a_3 =0.
$$
Equating the coefficients of the highest powers of $u$ and $v$ in $Q$ to zero, we get 
\begin{equation}\label{hig}
\sum_{i=1}^{k} n_i\, (a_1-\kappa_i b_3)=0, \qquad  
\sum_{i=1}^{k} n_i\, \left(b_1- \frac{a_3}{\kappa_i}\right)=0.   
\end{equation}
If $k>3$, then all coefficients of the cubic polynomial are equal to zero. Since $\sum n_i \ne 0$, it follows from (\ref{hig})
that $a_1=b_1=0$ and therefore all coefficients of the system 
(\ref{absys}) are equal to zero. \quad $\square$

\rem\label{rem2} Analogously, one can prove that if a system 
\begin{equation}\label{twod}
\left\{
\begin{array}{lcl}
u_t &=& P_1(u,v), \\[2mm] 
v_t &=&  P_2(u,v),
\end{array}
\right.
\end{equation}
where $P_i$ are homogeneous polynomials of degree $d$, has a first integral (\ref{fint}), then $k\le d+1.$

In the following we look into the three cases of 3, 2 and 1 distinct roots $\kappa_i$.

{\bf Case of 3 roots. } Consider the most non-degenerate case of three   distinct roots $\kappa_i$. Since system (\ref{absys})
is defined up to the group of linear transformations (\ref{invert}),  
we may reduce the integral to   

\begin{equation} \label{3roots}
I=u^{k_1}(u-v)^{k_2} v^{k_3},
\end{equation} 
where $k_i$ are natural numbers which are defined up to permutations. Without loss of generality we will assume that $$k_1\le k_2\le k_3$$
and that $k_1,k_2,k_3$ have not any non-trivial common divisor.

\lem\,\,  A system (\ref{absys}) has an integral (\ref{3roots}) iff 
up to a scaling $u\to \mu u,\, v \to \mu v$ it has the following form:
\begin{equation}\label{sys3root}
\left\{
\begin{array}{lcl}
u_t &=&  -k_3 \, u^2 + (k_3+k_2) \,u v  \\[2mm]
v_t &=& - k_1 \, v^2 + (k_1+k_2) \, u v.
\end{array}
\right.
\end{equation}

Let us apply now the  Kowalevski-Lyapunov integrability test to system (\ref{sys3root}). A straightforward computation leads to:

\theo \
The Kowalevski exponents for (\ref{sys3root}) have the form
\begin{equation}\label{nk}
s_1=\frac{k_2+k_3}{k_1}, \qquad s_2=\frac{k_3+k_1}{k_2}, \qquad s_3=\frac{k_1+k_2}{k_3}.
\end{equation}

The  system passes through the Kowalevski-Lyapunov test iff these numbers are natural.

\lem \label{lem3} There exist only three sets $k_1, k_2, k_3$ for which $s_i$ are natural numbers. They are:
\begin{itemize}
\item [\ ]Case 1. \quad $k_1=k_2=k_3=1$;
\item [\ ]Case 2. \quad $k_1=k_3=1, \quad k_2=2$;
\item [\ ]Case 3. \quad $k_1=1,\quad k_2=2,\quad k_3=3$.
\end{itemize}

\rem In Case 1 the  degrees of symmetries described by Lemma \ref{lem1} are $2+3 n, \, n \in \N$, for Case 2 they are $2+4 n$ and in Case 3 we have 
the degrees $2+6 n$.

\rem \label{rem5} Remark \ref{rem2}, applying together with the Kowalevski-Lyapunov test, allows one to describe integrable scalar homogeneous systems of the form (\ref{twod}) with $d>2$ as well. For example, the cubic system
\begin{equation}\label{4root}
\left\{
\begin{array}{lcl}
u_t &=&  -u^3\, k_3 + \nu \,u v^2 \,(k_2 + k_3 + k_4) + u^2 v \,(k_2 + k_3 - \nu k_3 - \nu k_4)  \\[2mm]
v_t &=& -\nu \,v^3 \,k_1 + u v^2\, (-k_1 + \nu k_1 + \nu k_2 - k_4) + u^2 v \,(k_1 + k_2 + k_4),
\end{array}   
\right.
\end{equation}
where $\nu \ne 0,1,\infty,$
has the most general possible first integral 
$$
I=u^{k_1} (u - v)^{k_2} v^{k_3} (u - \nu \,v)^{k_4}.
$$
Kowalevski-Lyapunov test distinguishes the following   13  sets $\{k_1,k_2,k_3,k_4\}$:
$$\{1, 3, 8, 12\},\quad  \{1, 2, 6, 9\},\quad  \{2, 3, 10, 15\},\quad  \{1, 1, 4, 6\},\quad  \{1, 4, 5, 10\}, \quad \{1, 2, 3, 6\},$$ 
$$\quad  \{1, 1, 2, 4\},\quad  \{1, 2, 2, 5\},\quad  \{1, 1, 1, 3\},\quad  \{1, 3, 4, 4\},\quad  \{1, 1, 2, 2\},\quad  \{2, 3, 3, 4\},\quad  \{1, 1, 1, 1\},$$
which lead to integrable systems of the form (\ref{4root}).
  
{\bf Case of 2 roots. } In the case of two distinct roots $\kappa_i$ in (\ref{fint}) the integral always can be transformed to $I=u^{k_1} v^{k_3}.$ The corresponding system has the form
\begin{equation}\label{2root}
\left\{
\begin{array}{lcl}
u_t &=&  -b_2 q \, u^2 + a_2 \,u v  \\[2mm] 
v_t &=& - a_2 q^{-1}  \, v^2 + b_2 \, u v,
\end{array}   
\right.
\end{equation}
where $\displaystyle q=\frac{k_3}{k_1}.$ 
The Kowalevski-Lyapunov test gives rise to $q=1.$ This means that for integrable cases
$$
I = u\, v
$$
and degrees of symmetries described by Lemma \ref{lem1} are $2+2 n,  \, n \in \N$.

The constants $a_2, b_2$ in (\ref{2root}) can be normalized (by a scaling $u\to \mu u,\, v \to \nu v$ and by the involution $u \leftrightarrow v$) to $a_2=b_2=1$ or to $a_2=1,b_2=0.$  

{\bf Case of 1 root. } If the integral (\ref{fint}) has a single root, then without loss of generality we may put $$I=u .$$ The corresponding system  
\begin{equation}\label{1root}
\left\{
\begin{array}{lcl}
u_t &=& 0  \\[2mm] 
v_t &=& b_1 \, v^2 + b_2 \,  v u +b_3 \, u^2
\end{array} 
\right.
\end{equation} 
satisfies the  Kowalevski-Lyapunov test. In this case the degrees of symmetries are $2+n,  \, n \in \N.$

Using the transformations $v\to s_i v+s_2 u$, we may reduce the set of constants $b_i$ to one of the following:
\begin{itemize}
\item $b_1=b_2=1, \quad b_3=0$,
\item $b_1=1, \quad b_2=b_3=0$,
\item $b_1=b_3=0, \quad b_2=1$,
\item $b_1=b_2=0, \quad b_3=1$.
\end{itemize}

Thus we described all systems (\ref{absys}) such that each of them has a polynomial integral and satisfies the  Kowalevski-Lyapunov test.

\section{Non-commutative systems} \label{homnonabelian}

Consider non-commutative ODE systems of the form
\begin{equation}\label{geneq}
\frac{d u^{i}}{d t}=F^{i}({\bf u}), \qquad {\bf u}=(u^1,...,u^n),
\end{equation}
where $u_i$ do not commute with each other and $F^{i}$ are (non-commutative)
polynomials with constant scalar coefficients. As usual, a symmetry
of (\ref{geneq}) is defined as an equation
\begin{equation}\label{gensym}
\frac{d u^{i}}{d \tau}=G^{i}({\bf u}), 
\end{equation}
compatible with (\ref{geneq}). 

More rigorously, we consider the free associative
algebra ${\mathfrak A}$ with generators $u^1,...,u^n$ over $\C$ and derivations of this algebra.

\defi \label{defi7} 
A linear map $d: {\mathfrak A}\to {\mathfrak A}$ is called derivation if it
satisfies the Leibniz rule: $d (u v)= d(u) v + u d(v).$

If we fix $d(u^i)=F^i({\bf u}),$  
then $d(z)$ is uniquely defined for any element $z\in {\mathfrak A}$ by
the Leibniz rule. It is clear that the polynomials $F^i$ can be taken
arbitrarily. Instead of the dynamical system (\ref{geneq}) we consider
the derivation $D_t: {\mathfrak A}\to {\mathfrak A}$ such that 
$D_t(u^i)=F^i.$
Compatibility of (\ref{geneq}) and (\ref{gensym}) means that the
corresponding derivations $D_t$ and $D_{\tau}$ commute: $D_t
D_{\tau}-D_{\tau} D_t=0.$

A non-commutative system (\ref{geneq}) will be regarded as  {\it integrable} if it
possesses infinitely many linearly independent symmetries.

\exam \label{exam1} \cite{miksokcmp} The following non-commutative system 
\begin{equation}\label{uv}
\left\{
\begin{array}{lcl}
u_t &=& v^2  \\[2mm] 
v_t &=& u^2
\end{array} 
\right.
\end{equation} 
is integrable. Substituting $m\times m$ matrices for $u$ and $v,$ we
obtain a matrix system integrable by the inverse scattering method. Many further reductions of (\ref{uv}) are
possible. For instance, F. Calogero has observed that in the matrix
case the functions $x_i=\lambda_i^{1/2}$, where the $\lambda_i$ are
eigen-values of the matrix $u-v,$ satisfy the following integrable
system:
$$
x_i^{''}=-x_i^5+\sum_{j\ne i} (x_i-x_j)^{-3}+(x_i+x_j)^{-3}.
$$ 

Thus integrable non-commutative systems are fundamental  models, which generate
plenty of different integrable finite-dimensional ODE systems by reductions.

The system (\ref{uv}) belongs to the class of homogeneous quadratic
non-commutative systems of the form
\begin{equation}\label{quadN}
(u^i)_t = \sum_{j,k} C^i_{jk}\, u^j \,u^k, \qquad i,j,k=1,\dots, n,
\end{equation}
where $C^i_{jk}$ are (complex) constants. 
 
The class (\ref{quadN}) is invariant with respect to the group $GL_n$
of linear transformations
$$
\hat u^i = \sum_{j} s^i_{j}\, u^j.
$$

In this paper we consider systems of the form (\ref{equ}), which belong to class (\ref{quadN}) with $n=2$.  Some rational non-commutative systems with $n=2$ were considered in \cite{WoEf,sqlinsolve}. For non-commutative PDEs see \cite{OlvSok98}.

Define a $\C$-linear involution
$\star$ on ${\mathfrak A}$ by the formulas
\begin{equation}\label{star1}
u^\star=u\, ,\qquad v^\star=v\, ,\qquad (a\, b)^\star=
b^\star \, a^\star \, ,\qquad a,b\in {\mathfrak A}.
\end{equation}
In the matrix case the transposition is such an involution.

\defi Two systems (\ref{equ}) related to each other by a linear
transformation of the form (\ref{invert})
and by involution (\ref{star1}) are called {\it equivalent}.
Under transformations  (\ref{invert}) and  (\ref{star1}) symmetries go to symmetries and integrable equations remain to be integrable. 

\defi \label{trian} A system that is equivalent to a system (\ref{equ}) with
$\beta_2=\beta_3=\beta_4=0$ is called {\it triangular}. If
in addition $\beta_1=0,$ then the system is called {\it 
strongly triangular}.

Sometimes it is useful to extend the free associative algebra  ${\mathfrak A}$ with generators $u$ and $v$ by new symbols $u^{-1}$ and $v^{-1}$ such that
$u u^{-1} = u^{-1} u = v v^{-1} = v^{-1} v = \mbox{1}.$ We will call the elements of the extended algebra {\it non-commutative Laurent polynomials}. 
For some classes of  polynomial systems (\ref{equ}) the set of admissible transformations (\ref{invert}), 
 (\ref{star1}) can be extended by special invertible Laurent transformations.  For example, consider systems of the form
\begin{equation}\label{shor}
\left\{
\begin{array}{lcl}
u_t &=&  -p\, u^2 + q \,u v \\[2mm] 
v_t &=& -a \, v^2 + b \, u v + c\, v u.
\end{array}   
\right.
\end{equation}
It can be easily verified that the composition of the transformation 
$$
u = \bar u, \qquad v=\bar u^{-1} \bar v \bar u
$$
and the involution  (\ref{star1}) maps (\ref{shor})
to 
\begin{equation}\label{newshor}
\left\{
\begin{array}{lcl}
u_t &=&  -p\, u^2 + q \,u v \\[2mm]
v_t &=& -a \, v^2 + (c+p) \, u v  + (b-p)\, v u.
\end{array}   
\right.
\end{equation}
Thus we have an involution $\tau: (\ref{shor}) \to (\ref{newshor})$ on the set of systems of the form (\ref{shor}).

\rem For any system of the form (\ref{shor}) with $q \ne 0$ we can express $v$ via $u_t, u$ and $u^{-1}$ from the first equation. Substituting this expression to the second equation, we obtain a second order non-commutative  Laurent equation for $u$.
 
\subsection{Known results}\label{Sec3.1}
 
Some experiments with non-triangular systems (\ref{equ}) having
symmetries have been done in \cite{miksokcmp}. 
One of the results is:

\theo Any non-triangular system (\ref{equ}) possessing
 a non-zero symmetry of the form
$$ 
 \left\{
\begin{array}{lcl}
u_{\tau}  &=&\gamma _1 u\, u\, u  + \gamma _2 u\, u\, v +
\gamma _3 u\, v\, u + \gamma _4 v\, u\, u \, +   
\gamma _5 u\, v\, v  + \gamma _6 v\, u\, v +
\gamma _7 v\, v\, u + \gamma _8 v\, v\, v ,\\[3mm]
v_{\tau}  &=& \delta _1 u\, u\, u  + \delta _2 u\, u\, v + \delta _3
u\, v\, u + \delta _4 v\, u\, u \, +  
\delta _5 u\, v\, v  + \delta _6 v\, u\, v +
\delta _7 v\, v\, u + \delta _8 v\, v\, v
\end{array}   
\right.
$$ 
is equivalent to one of the following:
\begin{equation} 
 \left\{
\begin{array}{lcl} 
u_t&=&u\, u - u\, v  \\[2mm]
v_t&=&v\, v - u\, v + v\, u
\end{array}  ,
\right. \label{casea}
\end{equation}
\[\left\{
\begin{array}{lcl} 
u_t&=&u\, v  \\[2mm]
v_t&=&v\, u
\end{array}  ,
\right.\]
\[ \left\{
\begin{array}{lcl} 
u_t&=&u\, u - u\, v  \\[2mm]
v_t&=&v\, v - u\, v
\end{array}  , 
\right.\]
\[ \left\{
\begin{array}{lcl} 
u_t&=& - u\, v  \\[2mm]
v_t&=&v\, v + u\, v - v\, u
\end{array}  , 
\right.\]
\[ \left\{
\begin{array}{lcl} 
u_t&=&u\, v - v\, u  \\[2mm]
v_t&=&u\, u + u\, v - v\, u
\end{array}  ,
\right.\]
\[ \left\{
\begin{array}{lcl} 
u_t&=&v\, v  \\[2mm]
v_t&=&u\, u
\end{array}  .
\right.\]

\rem It is a remarkable fact that a requirement of the existence of just
one cubic symmetry selects a finite list of equations without free
parameters (or more precisely, all possible parameters can be
removed by linear transformations (\ref{invert}) ).

\qquad  The following five non--equivalent systems with quartic symmetries were found in \cite{miksokcmp}:

\[ \left\{
\begin{array}{lcl}   
u_t&=&- u\, v \\[2mm]
v_t&=&v\, v + u\, v
\end{array}  ,
\right.\]
\[ \left\{
\begin{array}{lcl} 
u_t&=&-v\, u  \\[2mm]
v_t&=&v\, v + u\,v
\end{array}  ,
\right.\]
\[ \left\{
\begin{array}{lcl} 
u_t&=& u \, u - 2 v\, u \\[2mm]
v_t&=&v\, v  - 2 v\, u
\end{array}  ,
\right.\]
\[ \left\{
\begin{array}{lcl} 
u_t&=&u\, u - u\, v - 2 v\,u \\[2mm]
v_t&=&v\, v -v \, u - 2 u\, v
\end{array}  ,
\right.\]
\begin{equation} \left\{
\begin{array}{lcl} 
u_t&=&u\, u - 2 u\, v \\[2mm]
v_t&=&v\, v + 4  v\, u
\end{array}  . 
\right. \label{caseb}
\end{equation}
Using computer algebra system CRACK \cite{CRACK}, we verified that it is a complete list of non--triangular systems that have no cubic but quartic symmetries.
 
Possibly systems (\ref{casea})  and (\ref{caseb}) have only one symmetry while the other 9 have infinitely many.

Attempts to describe  systems  (\ref{equ}) with fifth degree symmetries by a straightforward computation look rather hopeless. One of  the reasons is that the coefficients of (\ref{equ}) turn out to be related by algebraic relations. Even if they  can be resolved, the coefficients very often become algebraic numbers.

\exam The non-commutative system
$$
\left\{
\begin{array}{lcl}
u_t &=& 11 \sqrt{7}\, u\, u -  7 \sqrt{7} \, v\, v \\ [2mm]
v_t &=& -4 \sqrt{7}\,v\,u  - 4 \sqrt{7}\,u\,v + 30 \, u\,u
\end{array}   
\right.
$$
has a symmetry of fifth degree.  Notice that in the commutative case this system is not integrable: it has no polynomial first integrals and does not satisfy the Painlev\'e test. 

Our crucial requirement is that the commutative limit of integrable non-commutative systems should be integrable in the sense of Section 2. It is true for all systems found in \cite{miksokcmp} except for systems (\ref{casea})  and (\ref{caseb}). So our requirement discards from the lists of paper \cite{miksokcmp} systems, which do not have an infinite hierarchy of symmetries.

\subsection{Non-commutative generalizations in the case of three roots}

The general ansatz for the non-commutative generalizations (see Introduction for the definition) is given by 
\begin{equation}\label{3rootN}
\left\{
\begin{array}{lcl}
u_t &=&  -k_3 \, u^2 + (k_2+k_3) \,u v  + \alpha (u v - v u)\\ [2mm]
v_t &=& - k_1 \, v^2 + (k_1+k_2) \, v u + \beta (v u - u v).
\end{array}   
\right.
\end{equation}
 We are going to find the parameters $\alpha$ and $\beta$ such that this non-commutative generalization is integrable in the sense described in Introduction. 
 
Involution (\ref{star1}) transforms $\alpha$ and $\beta$ in (\ref{3rootN}) as follows:
$$
\alpha \to -\alpha - k_2 - k_3, \qquad  \beta \to -\beta - k_2 - k_1. 
$$

Integrable commutative systems of the form (\ref{sys3root}) were described in Lemma \ref{lem3}.
In Case 1 we have $k_1=k_2=k_3=1.$ According to  Lemma \ref{lem1} and the assumption {\bf 2}  (see Introduction), any integrable non-commutative generalization (\ref{3rootN}) has to have 
a symmetry of degree 5 whose commutative limit is given by
\begin{equation}\label{Case1sym}
\left\{
\begin{array}{lcl}
u_{\tau} &=& \Big(-  \, u^2 + 2 \,u v  \Big) \,u \,v \,(u-v) \\ [2mm]
v_{\tau} &=& \Big(- \, v^2 + 2 \, v u    \Big) \,u\, v \,(u-v) .
\end{array}   
\right.
\end{equation}

\theo \label{theo3}  In the case $k_1=k_2=k_3=1$ there exist only 5 non-equivalent non-commutative generalizations that have a fifth degree symmetry with commutative limit (\ref{Case1sym}). They correspond to the following pairs $\alpha, \beta$ in (\ref{3rootN}):
\begin{itemize}
\item [\bf 1.] \quad  $\alpha = -1, \quad  \beta = -1$,
\item [\bf 2.] \quad  $\alpha = \ \ 0, \quad  \beta = -1$,
\item [\bf 3.] \quad  $\alpha = \ \ 0, \quad  \beta  = - 2$,
\item [\bf 4.] \quad  $\alpha = \ \ 0, \quad  \beta = \ \ 0$,
\item [\bf 5.] \quad  $\alpha = \ \ 0, \quad  \beta = -3$.
\end{itemize}
{\bf Outline of the proof.} We consider an ansatz with undetermined coefficients for a non-commutative symmetry of degree 5 such that its commutative limit coincides with (\ref{Case1sym}). The compatibility of (\ref{3rootN}) and the symmetry lead to an overdetermined linear algebraic system for the symmetry coefficients. The coefficients of this system depend on the parameters $\alpha$ and $\beta$. A tedious but automated analysis of the linear system gives us all pairs $\alpha, \beta$ for which the system has a non-trivial solution. The corresponding calculations were done by the computer algebra system CRACK. In the last step pairs were selected that correspond to non-equivalent systems. \quad $\square$

\rem The system of case {\bf 1} is equivalent to the system from Example 1. It has also symmetries of degrees 3 and 4 but their commutative limits are trivial. Systems of cases {\bf 2} and {\bf 3} are equivalent to some systems found in \cite{miksokcmp}. Systems of cases {\bf 4} and {\bf 5} are new.

In the cases ${\bf 2} -{\bf 5}$ of Theorem \ref{theo3} the  systems have the form (\ref{shor}). Applying the Laurent involution  $\tau: (\ref{shor}) \to (\ref{newshor})$ which corresponds to $\beta \to -\beta-3,$ we see that the cases ${\bf 2},{\bf 3}$ and ${\bf 4}, {\bf 5}$ are dual with respect to $\tau$.  \vspace{12pt}

Consider now Case 2 of Lemma \ref{lem3}: $k_1=k_3=1, \, k_2=2$. The integrable non-commutative generalizations should have symmetries of degree 6 with the prescribed commutative limit defined by Lemma \ref{lem1}.

\theo \label{theo4}  In the case $k_1=k_3=1, \, k_2=2$ there exist only 4  non-equivalent non-commutative generalizations that have the symmetry of degree six. They correspond to:
\begin{itemize}
\item [\bf 1.] \quad  $\alpha = -1, \quad  \beta = -1$,
\item [\bf 2.] \quad  $\alpha = \ \ 0, \quad  \beta = -2$,
\item [\bf 3.] \quad  $\alpha = \ \ 0, \quad  \beta = \ \ 0$,
\item [\bf 4.] \quad  $\alpha = \ \ 0, \quad  \beta = -4$.
\end{itemize}

\rem  The systems {\bf 2,\,3} and {\bf 4} are new.

Under the Laurent involution $\tau$: (\ref{shor}) $\to$ (\ref{newshor}) the parameter $\beta$ is changing as $\beta \to -\beta-4.$ So, the cases {\bf 3} and {\bf 4} are dual and the case {\bf 2} is self-dual.\vspace{12pt}

In Case 3 of Lemma \ref{lem3}: $k_1=1,\, k_2=2, \, k_3=3$ any integrable non-commutative generalization should have a symmetry of degree 8.

\theo \label{theo5}  In the case  $k_1=1,\, k_2=2, \, k_3=3$ there exist only 5  non-equivalent non-commutative generalizations with the symmetry of degree 8. They correspond to:
\begin{itemize}
\item [\bf 1.] \quad  $\alpha = -2, \quad  \beta = \ \ 0$,
\item [\bf 2.] \quad  $\alpha = -4, \quad  \beta = \ \ 0$,
\item [\bf 3.] \quad  $\alpha =  -6, \quad    \beta = \ \ 0$,
\item [\bf 4.] \quad  $\alpha = \ \ 0, \quad \beta = -6$,
\item [\bf 5.] \quad  $\alpha = \ \ 0, \quad  \beta = \ \ 0$.
\end{itemize}

\rem  All these non-commutative systems  are new.

For systems with $\alpha=0$ the involution $\tau$ corresponds to $\beta \to -\beta-6$ and therefore, systems {\bf 4} and {\bf 5} are dual. For systems with $\beta=0$ we may apply the Laurent transformation 
$\pi: v=\bar v,\, u=\bar v^{-1} \bar u \bar v.$ The composition of this transformation and the involution (\ref{star1}) leads to the system with $\bar \beta=0,\, \bar \alpha=-\alpha -6.$ Hence, the systems of cases {\bf 1} and {\bf 2} of Theorem \ref{theo5} are dual and the system of case {\bf 3} is self-dual.
  
\rem In the cases of Theorems \ref{theo3} and \ref{theo4} we do not consider the Laurent involution $\pi$ for systems with  $\beta=0$. The reason is that the systems with $k_1=k_2=k_3=1$  and with $k_1=k_3=1, \, k_2=2$ are invariant with respect to $u \leftrightarrow v$, $\alpha \leftrightarrow \beta.$

\subsection{Non-commutative generalizations in the cases of two and one roots}

The possible non-commutative generalizations of (\ref{2root}) with $q=1$ have the form
\begin{equation}\label{2rootN}
\left\{
\begin{array}{lcl}
u_t &=&  -b_2 \, u^2 + a_2 \,u v  + \alpha (u v - v u)\\[2mm]
v_t &=& - a_2 \, v^2 + b_2 \, v u + \beta (v u - u v).
\end{array}   
\right.
\end{equation}
The degree of the prescribed symmetry is 4 and therefore all non-commutative generalizations have to be equivalent to some systems from 
\cite{miksokcmp} (see Section \ref{Sec3.1}).

The involution (\ref{star1}) corresponds to $\bar \alpha = -\alpha-a_2,\, \bar \beta = -\beta-b_2$.

\theo \label{theo6} There exist only 4 non-triangular non-equivalent non-commutative systems (\ref{2rootN}) with a fourth degree symmetry. They correspond to
\begin{itemize}
\item [\bf 1.] \quad  $a_2=b_2=1$, \qquad \quad $\alpha = 0, \qquad\,  \beta = \ \ 0$,
\item [\bf 2.] \quad   $a_2=b_2=1$, \qquad \quad $\alpha = 0, \qquad \, \beta = -2$,
\item [\bf 3.] \quad   $a_2=b_2=1$, \qquad \quad $\alpha = 0, \qquad \, \beta = -1$,
\item [\bf 4.] \quad   $a_2=1,\,\, b_2=0$,\qquad$\alpha = 0, \qquad \, \beta = \ \ 1$.
\end{itemize}

In this case the Laurent involution $\tau$ corresponds to $\bar \beta = -\beta - 2\, b_2$. Case {\bf 1} is dual to Case {\bf 2}. Case {\bf 3} is self-dual. In Case {\bf 4} we get $\bar \beta=-1$ but, using $u\to -u$ we may reduce it to $\beta=1.$ Therefore, Case {\bf 4} is also self-dual.  \vspace{12pt}

The non-commutative generalizations of (\ref{1root}) are given by 
\begin{equation}\label{1rootN}
\left\{
\begin{array}{lcl}
u_t &=&  \alpha (u v - v u)\\[2mm]
v_t &=& b_1 \, v^2 + b_2 \, v u + b_3 u^2+\beta (v u - u v).
\end{array}   
\right.
\end{equation}
The symmetry degree is three.

\theo \label{theo7}  There exist only 2 integrable non-triangular non-equivalent systems (\ref{1rootN}) with a symmetry of third degree. They correspond to
\begin{itemize}
\item [\bf 1.] \quad  $b_1=b_2=1,\quad b_3=0$, \qquad \quad $\alpha = 1, \qquad  \beta = 0$,
\item [\bf 2.] \quad    $b_1=b_2=0,\quad b_3=1$, \qquad \quad $\alpha = 1, \qquad  \beta = 0$.
\end{itemize} \vspace*{9pt}

{\bf Conjecture.} Each one of the systems described in Theorems \ref{theo3} - \ref{theo7} possesses an infinite hierarchy of symmetries. Any non-triangular system of the form (\ref{equ}), possessing such a hierarchy, is equivalent to one of systems from Theorems \ref{theo3} - \ref{theo7}.

We have verified that all these systems have more than one symmetry of degree not greater than 16 apart from systems (\ref{casea})  and (\ref{caseb}) from Section \ref{Sec3.1}. 

\section{Integrable inhomogeneous generalizations} \label{inhomgen}

In this section we find all integrable inhomogeneous generalizations of the form (\ref{inhom}) for each system (\ref{equ}) from Theorems 3--7. It turns out that commutative limits of these inhomogeneous systems have first integrals, which are inhomogeneous generalizations of the corresponding homogeneous integrals from Section \ref{homabelian}.

 We call two inhomogeneous generalizations {\it equivalent} if they are related by a shift 
\begin{equation}\label{shift}
u \to u + c_1 \, {\rm I}, \qquad  v \to v + c_2 \, {\rm I},
\end{equation}
where $c_i$ are constants. 

\subsection{The case of Theorem 3}

We consider the systems from cases {\bf 1} - {\bf 5} of Theorem 3.  By a shift of the form  (\ref{shift}) we reduce $\gamma_2$ and $\gamma_4$ in (\ref{inhom}) to zero.

\prop\label{prop2} For each of cases {\bf 1} - {\bf 5} of Theorem 3 there exists an inhomogeneous generalization with a fifth degree inhomogeneous symmetry iff $\gamma_5 = - \gamma_1$.

The commutative limits of all these generalizations have the same polynomial cubic integral 
$$ I = v (v-u) u + \gamma_1 v u + \gamma_3 v - \gamma_6 u, $$
which is an inhomogeneous generalization of the integral (\ref{3roots}).
\subsection{The case of Theorem 4}

In the cases {\bf 1} - {\bf 4} of Theorem 4 we also reduce $\gamma_2$ and $\gamma_4$ in (\ref{inhom}) to zero by a shift of the form  (\ref{shift}).  

\prop For each of cases {\bf 1} - {\bf 4} of Theorem 4 there exists inhomogeneous generalization with an inhomogeneous symmetry of degree six iff $\gamma_5 = - \gamma_1$ and $\gamma_6 = \gamma_3$. 

The commutative limits of all these generalizations have the same polynomial integral of fourth degree
$$ I = 2 u (u - v)^2 v + 4 \gamma_5 \,u (u - v) v + \gamma_3 \,(u - v)^2 + 
 2 \gamma_5^2\, u v + 2 \gamma_3 \gamma_5 \,(u - v) .$$

\subsection{The case of Theorem 5}

We consider the systems from cases {\bf 1} - {\bf 5} of Theorem 5. All of them have  homogeneous symmetries of degree 8.  By a shift of the form (\ref{shift})  we reduce $\gamma_2$ and $\gamma_4$ to zero.

\prop Each of cases {\bf 1} - {\bf 5} of Theorem 5 has two  different inhomogeneous generalizations with inhomogeneous symmetry of eight degree:
\begin{itemize}
\item [\bf a.] \qquad $\gamma_1 = -3 \,\gamma_5, \quad \gamma_3 = \gamma_6 = 0$,
\item [\bf b.] \qquad $\gamma_1 = \gamma_5 = \gamma_6 = 0$.
\end{itemize}
All commutative limits for systems from the item {\bf a} have the same first integral
$$ I = u (u - v)^2 v^3 + 2 \gamma_5 \,u (u - v) v^3 + \gamma_5^2\, u v^3,$$
while the commutative limits for the item {\bf b} have the integral 
$$ I = 4 u (v-u)^2 v^3   + \gamma_3\, (v-u) (v+3u) v^2 + \gamma_3^2\,v^2. $$

\subsection{The case of Theorem 6}

We consider the systems from cases {\bf 1} - {\bf 4} of Theorem 6. 

\prop In each of cases {\bf 1} - {\bf 3} of Theorem 6 by shift  (\ref{shift}) we reduce $\gamma_2$ and $\gamma_4$ to zero. Inhomogeneous generalizations, which have an inhomogeneous symmetry of degree four, exist iff
$$\gamma_1 = - \gamma_5, \qquad \gamma_3 = \gamma_6 =0.$$
The common first integral for all commutative limits is 
$$I = u \,v.$$ 

\prop In case {\bf 4} of Theorem 6 we can reduce $\gamma_1$ and $\gamma_2$ to zero by a shift.
For the integrable inhomogeneous generalizations with a fourth degree symmetry we get
$$\gamma_3 = \gamma_4 = \gamma_5 = \gamma_6 = 0.$$
Thus, non-trivial integrable inhomogeneous generalizations do not exist.
 
\subsection{The case of Theorem 7}

\prop In case {\bf 1} of Theorem 7 by shift  (\ref{shift}) we can reduce $\gamma_4$ and $\gamma_5$ to zero. 
Inhomogeneous generalization with a third degree symmetry exists iff
$$\gamma_1 = \gamma_2 = \gamma_3=0.$$
The integral for the commutative limit is $I=u.$

\prop In case {\bf 2} of Theorem 7 we reduce $\gamma_4$ to zero by a shift.
 For the integrable inhomogeneous generalizations with a symmetry of degree 3 we get
$$\gamma_1 = \gamma_2 = \gamma_3 = \gamma_5 = \gamma_6 = 0.$$
Thus, non-trivial integrable inhomogeneous generalizations do not exist.

\section{On integrability of non-commutative systems in the matrix case}\label{section5}

All non-commutative systems found in Sections 3 and 4 become systems of ODEs with $2 m^2$ variables if we replace $u$ and $v$ by $m\times m$-matrices. Probably all these ODE systems are integrable but it should be proved separately for each case. Different types of integrability arise here. By analysing the structure of independent first integrals and symmetries, one can try to predict the integrability procedure. 

For some systems it is possible to find a general solution in an explicit form.   For example, the first system from Theorem \ref{theo7} can be reduced by the Laurent triangular transformation $u\to v^{-1} u v$ to a triangular system 
$$
\left\{
\begin{array}{lcl}
u_t &=& 0\\  
v_t &=&  \, v^2 +  \, u v.
\end{array}   
\right.
$$
Let $u = c$, where $c$ is a non-degenerate constant matrix. Then we get the following linear matrix equation
$$
w_t = - w\, c - {\bf 1}
$$
for $w = v^{-1}.$ The latter equation has the general solution
$$
w = - c^{-1} + c_1 \exp{(- c \,t)},
$$
where $c_1$ is an arbitrary constant matrix. The case of degenerate matrix $c$ also can easily be integrated in quadratures. 

Some of systems can be solved or linearized by the AKS factorization method \cite{AKS, golsok4}. For instance, the second system of Theorem \ref{theo7} implies
\begin{equation} \label{vv}
v_{tt} = [v_t,\, v].
\end{equation}
In the matrix case this equation can be reduced to a linear equation in the following way. If $Y$ is a matrix solution of the linear equation
$$
Y_t = Y (c_1 t+c_2),
$$
where $c_i$ are arbitrary constant matrices, then $v=-Y_t Y^{-1}$ is a general solution of (\ref{vv}).

Several systems can probably be solved by the inverse scattering method. To do that we have to find Lax pairs for them. 
In the next section  we present a Lax pair for the inhomogeneous generalization of the system from Example 1.

\subsection{Integrable matrix generalization of a flow on an elliptic curve}\label{sub6.1}

Let us consider the  system of ODEs (\ref{inuv}).  
Its homogeneous limit  is the system from Example 1, which is equivalent to the system described by Case 1 of Theorem 3.

In the case $m=1$ we have a system of two ODEs, which can be written in the Hamiltonian form 
$$
u_t = - \frac{\partial H}{\partial v}, \qquad v_t =  \frac{\partial H}{\partial u}
$$
with the Hamiltonian 
$$
H = \frac{1}{3} u^3 - \frac{1}{3} v^3 - c u v + b u - a v.
$$
For generic $a,b,c$ the relation $H = const$ is an elliptic curve   and (\ref{inuv}) describes the motion of  its point. 

In the case of arbitrary $m$ the system (\ref{inuv}) remains to be Hamiltonian with the Hamiltonian
\begin{equation}\label{HAMmat}
H = {\rm tr}\, \left( \frac{1}{3} u^3 - \frac{1}{3} v^3 - c u v + b u - a v \right)
\end{equation} 
and non-abelian constant Poisson bracket \cite{odrubsok}.

We believe that the integrable non-commutative system (\ref{inuv}) could be one of the  stones on which a theory of non-abelian elliptic functions can be built.

\subsection{Lax representation}

In the homogeneous matrix case (\ref{uv})  the following Lax $(L,A)$-pair 
\begin{equation}\label{HomLax}\begin{array}{c}
L = \left(\begin{array}{ccc} 1 & 0 & 0\\ 0 & \varepsilon & 0\\0 & 0 & \varepsilon^2 \end{array}\right) \, \lambda + 
\left(\begin{array}{ccc} 0 & 3 \varepsilon u & 3 v \\ v & 0 & ( \varepsilon -1) u\\u& (2 \varepsilon+1) v & 0 \end{array}\right), \\ \\
\displaystyle A = - \frac{1}{3} \left(\begin{array}{ccc} \varepsilon^2 & 0 & 0\\ 0 & \varepsilon & 0\\ 0 & 0 & 1 \end{array}\right) \, \lambda + 
\frac{1}{3} \left(\begin{array}{ccc} 0 & 3 \varepsilon^2 u & 3 v\\ \varepsilon v & 0 & ( \varepsilon +2) u\\u& (1-\varepsilon) v&0 \end{array}\right),
\end{array}
\end{equation}
where 
\begin{equation}\label{eps}
\varepsilon^2+\varepsilon+1=0,
\end{equation}
can be derived from Section 3.1 of \cite{miksokcmp} and from the Lax pair 
$$
L = \lambda\,C + M, \qquad A = \frac{1}{\lambda} \, M^2
$$
of the non-abelian Manakov equation  \cite{man}
$$M_t = [M^2, \, C].$$ 
 In this equation $M(t)$ is an unknown matrix of arbitrary size and $C$ is a given constant matrix. 

\prop The Lax equation 
\begin{equation} \label{Lax}
\bar L_t = [A,\,\bar L],
\end{equation}
where
\begin{equation}\label{InHomLax} 
\bar L = \lambda\, L + \lambda\, c\, P + a\,Q + b\, R,
\end{equation}
\medskip
{\small
$$
P=\left(\begin{array}{ccc}\varepsilon+2&0&0\\0&-2\varepsilon -1&0\\0&0&\varepsilon-1\end{array}\right),\, 
Q=\left(\begin{array}{ccc}0&3(\varepsilon+2)&0\\0&0&-3\\ \varepsilon -1&0&0\end{array}\right),\,
R=\left(\begin{array}{ccc}0&0&3(1-\varepsilon)\\2 \varepsilon+1&0&0\\0&-3\varepsilon&0\end{array}\right)
$$
}
and $L, A$ are given by (\ref{HomLax}), is equivalent to  system (\ref{inuv}).

\rem Formulas (\ref{Lax}) and (\ref{InHomLax}) define also the Lax representation for (\ref{inuv}) with $u$ and $v$ being elements of any associative algebra. 

\subsection{First integrals}

It is easy to see that all components of the matrix $uv - vu$ are first integrals for system (\ref{inuv}). Apart of these integrals there exists a sequence of integrals of the form  ${\rm tr}\,(P(u,v))$, where $P$ is a polynomial. We called them {\it trace first integrals}.  

It is clear that  ${\rm tr}\, M_n,$, where $ M_n=(vu-uv)^n,$ is a trace first integral   of degree $2n$.  
Non-trivial trace integrals come from the Lax representation.
Namely, it follows from (\ref{Lax}) that
\begin{equation}\label{trr}
\Big({\rm tr}\, (\bar L^k)\Big)_t = 0
\end{equation}
for any $k.$  Each expression ${\rm tr}\, (\bar L^k)$ is a polynomial in $\lambda$, whose 
coefficients are trace first integrals. All these integrals involve parameters $a, b, c.$ 

Moreover, replacing $\varepsilon^2$ by $-\varepsilon-1$ 
the resulting expressions are linear in $\varepsilon.$ Since $\varepsilon$ is any one of two solutions of the quadratic equation (\ref{eps}), we obtain two trace integrals through the 
coefficients of $\varepsilon^1, \varepsilon^0$. 

The integrals, which come from  (\ref{trr}) with $k=1,2$ are trivial. In the case $k=3$ the Hamiltonian (\ref{HAMmat}) 
arises. Relations (\ref{trr}) with $k=4,\dots,9$ produce the integrals ${\rm tr}\, (T_i)$ , where

\begin{eqnarray*}
\hspace*{-20pt}T_1&\hspace*{-10pt}=&v^6 - 6v^3u^3 + 6v^2uvu^2 - 2vuvuvu + u^6 + 6v^4uc - 6vu^4c + 6v^4a - 6vu^3a \\
 & & - 6v^3ub + 6u^4b + 9vuvuc^2 + 18v^2uac - 18vu^2bc - 18vuab + 9v^2a^2 + 9u^2b^2,\\[2mm]
\hspace*{-20pt}T_2&\hspace*{-10pt}=&v^5u^2 - 2v^4uvu + v^3uv^2u + 2vu^4vu - v^2u^5 - vu^3vu^2 + 3v^2uvu^2c - 3vuvuvuc\\
 & &+ 3v^3u^2a - 3v^2uvua - 3v^2u^3b + 3vu^2vub,\\[2mm]
\hspace*{-20pt}T_3&\hspace*{-10pt}=&v^5u^2vu - v^5uvu^2 - v^4u^2v^2u + v^4uv^2u^2 - v^3uv^2uvu + v^3uvuv^2u - v^2u^5vu\\
 & &+ v^2u^4vu^2 - v^2u^2vu^4 + v^2uvu^5 - vu^3vu^2vu + vu^3vuvu^2 + 3v^2u^2vuvuc \\
 & & - 3v^2uvuvu^2c + 3v^3u^2vua - 3v^3uvu^2a + 3v^2uvu^3b - 3v^2u^3vub,\\[2mm]
\hspace*{-20pt}T_4&\hspace*{-10pt}=&v^9 - 9v^6u^3 + 9v^5uvu^2 + 9v^4u^2v^2u - 9v^4uvuvu + 9v^3u^6 - 9v^3u^2v^3u + 9v^3uv^2uvu\\
 & &- 9v^2u^4vu^2 + 9v^2u^3vu^3 - 3v^2uv^2uv^2u - 9v^2uvu^5 + 9vu^4vuvu + 3vu^2vu^2vu^2\\
 & &- 9v^3u^3vuc - 9v^3u^2vu^2c - 9v^3uvu^3c - 27v^3uvubc - 9v^2u^3v^2uc + 9v^2u^2vuvuc\\
 & &+ 36v^2uvuvu^2c + 9vu^7c + 81vu^3b^2c - 18vuvuvuvuc - 9v^3u^2vua + 18v^3uvu^2a\\
 & &+ 18v^2u^2v^2ua - 9v^2uvuvua + 9vu^6a + 36v^3u^4b + 9v^2u^3vub - 18v^2u^2vu^2b\\
 & &- 18v^2uvu^3b + 9vu^2vuvub - 9u^7b - 27v^2u^4ac - 27v^2u^3a^2 + 27v^3u^2b^2\\
 & &+ 54vu^5bc - 27vu^4vuc^2 + 54vu^4ab - 27vu^3vuac - 27u^5b^2 + 81v^3ua^2c\\
 & &+ 81v^2uvuac^2 - 81v^2ua^2b - 81vu^2vubc^2 + 81vu^2ab^2 + 27vuvuvuc^3 \\
 & & - 162vuvuabc + 27u^3a^3 - 27u^3b^3 - 81vua^3c - 81va^4 + 81ua^3b.
\end{eqnarray*}

We verified that any trace integral of degree not greater than 9  is a linear combination of $H,$  $M_2,..., M_4$ and ${\rm tr}\, (T_i), \,\, i=1,\dots,4$.

According to Yu.\ Suris \cite{yusu} these trace integrals together with the components of the matrix $u v - v u$ provide a complete set of functionally independent integrals for the Hamiltonian system  (\ref{inuv}) in the cases of $2\times 2$ and $3\times 3$ - matrices.

\medskip

\section{Conclusion}

\qquad The ad hoc procedure for generalizing a given integrable polynomial model proposed in our paper opens a wide field for investigations. 
For example, the following problems look attractive:
\begin{itemize}
\item  [\bf 1.] It would be interesting to find Lax pairs for all systems described in Proposition  \ref{prop2}.

\item  [\bf 2.] Reasonably interesting is to find non-commutative generalizations of the cubic systems described in Remark \ref{rem5}.

\item  [\bf 3.] A separate promising problem is a non-commutative generalization of known integrable quadratic systems with more than two dependent variables. Our approach can be applied for that but the computations become more tedious. 

\item [\bf 4.](see Remark \ref{rem12}) An interesting problem is to find integrable  discretizations for non-commutative homogeneous systems from Theorems \ref{theo3} - \ref{theo7} and for their inhomogeneous generalizations.

\item  [\bf 5.](see Remark \ref{rem13})  The investigation of integrable Laurent deformation of non-commutative systems from Sections 3-4 looks very promising. 

\end{itemize}

\rem\label{rem12} System (\ref{inuv}) admits the following obvious discrete transformations, which change the values of the parameters $a, b, c$:
\begin{eqnarray*}
\hspace*{-20pt}R_1 &:&  \qquad \bar u = \varepsilon u,\qquad \bar v = \varepsilon^2 v, \qquad \bar a = \varepsilon a, \qquad \bar b = \varepsilon^2 b, \qquad \bar c = c ;\\ [3pt]
\hspace*{-20pt}R_2 &:&  \qquad \bar u = v,\qquad \bar v = u, \qquad \bar a = b, \qquad \bar b = a, \qquad \bar c = -c.  
\end{eqnarray*} 
In the scalar case V. Adler \cite{adler} found one more transformation
$$\hspace*{-20pt}R_3 :  \qquad \bar u = u + \frac{a-b}{v-u+c},\qquad \bar v =  v + \frac{a-b}{v-u+c}, \qquad \bar a = b, \qquad \bar b = a, \qquad \bar c = c. 
$$
It is easily verified that $R_1^3=R_2^2=R_3^2 = {\bf Id}$. The transformation $R_1 R_3 R_1 R_3$ does not change the parameters and  defines a non-trivial shift on the elliptic curve $H = {\rm const}.$ It can be verified that the transformation $R_3$ is applicable in the non-commutative case as well.

\rem\label{rem13} In the paper \cite{WoEf} the following integrable non-commutative  Laurent system 
$$
u_t = u v - u v^{-1} - v^{-1}, \qquad v_t = - v u + v u^{-1} + u^{-1},
$$
proposed by M. Kontsevich, was investigated. A Hamiltonian structure for it was discovered in \cite{artam}. This Laurent system can be regarded as a non-trivial deformation of the homogeneous system 
\begin{equation}\label{C1T7}
u_t = u v, \qquad v_t = - v u 
\end{equation}
by Laurent terms of smaller degree. System (\ref{C1T7}) is related to the system described by Case 1 of Theorem 7 through the transformation $u \to u + v, \,\, v \to - u.$

 {\bf Acknowledgments.} The authors are grateful to V. Adler,  M. Kontsevich, Yu. Suris  and   A. Odesskii for useful discussions.
The first author is grateful to IHES for hospitality and support. This work was carried out within the framework of state assignment 
No 0033-2019-0006 and of the State
Programme of the Ministry of Education and Science of the Russian
Federation, project No 1.12873.2018/12.1
and Canadian Network for Research and Innovation
in Machining Technology, Natural Sciences and Engineering Research Council of Canada (Grant No. RGPIN-2017-06330).

\section*{Conflict of interest}
On behalf of all authors, the corresponding author states that there is no conflict of interest.

\end{document}